\documentclass{vgtc}                          

\ifpdf
  \pdfoutput=1\relax                   
  \usepackage{graphicx}                
  \DeclareGraphicsExtensions{.pdf,.png,.jpg,.jpeg} 
\else
  \usepackage{graphicx}                
  \DeclareGraphicsExtensions{.eps}     
\fi%

\graphicspath{{figures/}{pictures/}{images/}{./}} 

\usepackage{microtype}                 
\PassOptionsToPackage{warn}{textcomp}  
\usepackage{textcomp}                  
\usepackage{mathptmx}                  
\usepackage{times}                     
\usepackage{cite}                      
\usepackage{tabu}                      
\usepackage{booktabs}                  
\usepackage{wrapfig}

\usepackage{colortbl}
\def\markup{1}
\if\markup1

\else

\fi

\usepackage{caption}

\onlineid{0}

\vgtccategory{Research}

\title{A Multi-scale Visual Analytics Approach for Exploring Biomedical Knowledge}

\author{Fahd Husain\thanks{e-mail: [fhusain, rgomez, ekuang, dsegura, acarolli, nliu, mcheung, yparis]@uncharted.software} %
\and Rosa Romero-Gómez %
\and Emily Kuang %
\and Dario Segura %
\and Adamo Carolli %
\and Lai Chung Liu %
\and Manfred Cheung %
\and Yohann Paris %
}
\affiliation{\scriptsize Uncharted Software Inc.}

\teaser{
  \centering
  \includegraphics[width=0.95\linewidth]{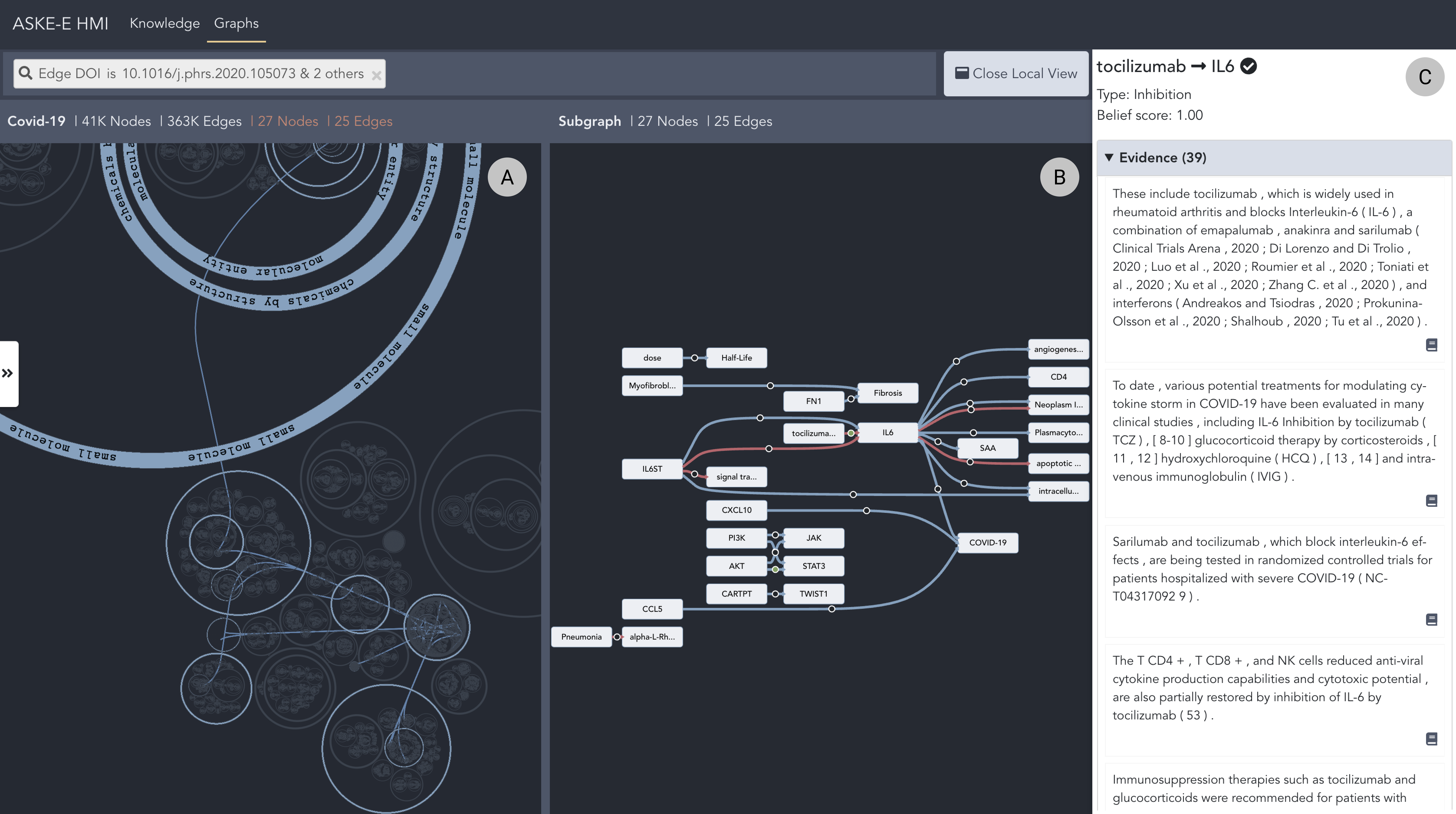}
  \caption{A biomedical researcher uses our system prototype to investigate potential drug treatments for SARS-CoV-2 using the COVID-19 biological graph that was automatically derived from literature. \textbf{A.} \textbf{The Global View} provides an overview of the biological graph with bundled edges and nodes organized hierarchically in a biomedical ontology. The results of searching for links from several articles using DOIs are highlighted. \textbf{B.} \textbf{The Local View} shows the highlighted results extracted as a node-link flow graph for further analysis. \textbf{C.} \textbf{The Drill-down Panel} displays the underlying evidence extracted from scientific articles for the inhibition relationship between \textit{tocilizumab} and \textit{IL6}. 
}
  \label{fig:teaser}
}

\abstract{
This paper describes an ongoing multi-scale visual analytics approach for exploring and analyzing biomedical knowledge at scale. We utilize global and local views, hierarchical and flow-based graph layouts, multi-faceted search, neighborhood recommendations, and document visualizations to help researchers interactively explore, query, and analyze biological graphs against the backdrop of biomedical knowledge. The generality of our approach - insofar as it requires only knowledge graphs linked to documents - means it can support a range of therapeutic use cases across different domains, from disease propagation to drug discovery.
Early interactions with domain experts support our approach for use cases with graphs with over 40,000 nodes and 350,000 edges.%
} 

\CCScatlist{
  \CCScatTwelve{Human-centered computing}{Visu\-al\-iza\-tion}{Visu\-al\-iza\-tion techniques}{Treemaps};
  \CCScatTwelve{Human-centered computing}{Visu\-al\-iza\-tion}{Visualization design and evaluation methods}{}
}

\begin{document}



\maketitle
\section{Introduction}
With over 1.5 million publications a year and more than 50 million peer-reviewed articles in existence, the rate and volume of novel scientific research remains overwhelming, having long surpassed our ability to fully understand and utilize what is known \cite{jinha2010article}. Crises like the COVID-19 pandemic only exacerbate this situation, with critical information dispersed across the papers published each day \cite{hechenbleikner_data_2020}. To stay up-to-date, scientists must manually review publications to internalize new knowledge. Therefore, their coverage of the field remains small, and at this scale and pace, contextualization of new knowledge and synthesis with prior knowledge is often impractical, biased, and error-prone. 
While all disciplines are subject to the effects of rapidly-changing knowledge, the cost in medical research can be measured in human lives.
 
In the field of biomedicine, it is already difficult and time-consuming to develop and validate hypotheses. Against the backdrop of all scientific knowledge, biomedical researchers start with a certain question, and turn to their prior domain knowledge and the mental models from their specific experiences \cite{janes_engineering_2017}. As many biological processes are inherently network phenomena, they are well suited for graph analysis \cite{sosa_literature-based_2019}. For example, when analyzing the effects of a drug on a disease, biomedical researchers must assemble (or update) biological graphs from literature, analyze their causal structure, extract relevant subgraphs, highlight novel or alternative pathways that could aid in drug discovery, flag potential viral mutations should new pathways appear, and distill all this analysis into a list of drug candidates for downstream trials. Similar graph analytic workflows are required across a range of biomedical use cases. However, the sheer size and dense connectivity of such graphs makes them hard to visualize and analyze in real-time. 

The primary research contribution of this paper is a multi-scale visual analytics approach that enables the investigation of biological graphs to facilitate a variety of use cases such as drug discovery, disease analysis, and identification of side-effects. This research is part of our ongoing effort for DARPA's Automating Scientific Knowledge (ASKE) program \cite{ASKE}.
Utilizing data from our ASKE collaborators \cite{INDRA, cosmos}, we developed a web-based prototype that visualizes 16 biological graphs against a corpus of 176,000 documents. 
We then conducted an expert evaluation in order to better understand how the system helps biomedical researchers in identifying drug-target interactions. 
Domain experts believe our approach to have unique value for knowledge exploration, and our initial iteration with users underscored that the prototype was easy to interact with and useful for navigating and analyzing large graphs against the backdrop of scientific knowledge. More generally, we believe our design approach is flexible enough to be used for knowledge discovery in other graph-based domains. 


\section{Related Work}

\subsection{Biological Graph Visualization}
Graph-based visualizations remain key to understanding biological graphs across a range of use cases \cite{gehlenborg_visualization_2010, pavlopoulos_survey_2008}. Most existing visualization approaches support static overviews of the entire graph \cite{kanehisa_kegg_2000, jeske_brenda_2019} and interactive views of focused subgraphs \cite{INDRA_pathwaymap, DialogueBio, SIGNOR}.
However, research has shown that the inherent multi-scale nature of biological graphs can only be fully appreciated when the entire range from local to global graph structures can be inspected continuously and interactively \cite{pirch_vrnetzer_2021}. 
Some tools such as Reactome \cite{sidiropoulos_reactome_2017} strike a balance by displaying overviews of biological graphs as well as detailed views of selected subgraphs or pathways. Yet, in the workflow, the overviews are replaced by selected pathways and thus global insight is decreased. By contrast, we propose a novel approach for scalable and performant graph analysis with coordinated and adjustable global and local views so the multi-scale character of biological graphs can be preserved and interrogated with multi-faceted search, interactive navigation, and progressive drill-down on demand.

\subsection{Large-Scale Scientific Knowledge Exploration}


Visualizations of large-scale scientific corpora can be broadly categorized into citation-linked graphs \cite{gates_nature_2019, eitan_connected_2021} and similarity clusters \cite{stasko_jigsaw_2013, kibardin_using_2020}.
Citation-linked graphs represent articles as nodes and edges as citations. Cluster visualizations typically apply dimensionality reduction to produce text embeddings that can be visualized as interactive 2-D scatterplots. For example, Open Knowledge Maps \cite{kraker_open_2016} displays clusters of PubMed \cite{PubMed} articles based on textual similarity.
More recent approaches visualize biomedical data repositories and scientific articles, using metadata for overview, organization, custom user-feeds and semantic querying \cite{lekschas_satori_2018, kraker_open_2016, kerren_biovis_2017}. 
  

Taking inspiration from the above work, we are building the capability for interactive exploration of hierarchical clusters of scientific knowledge, where nested sub-clusters organize related knowledge at finer resolution. To our knowledge, this visual and interactive approach is not yet present in the field at large.

\section{Design Process and Goals}

Our user-centered design process was based on periodic meetings over eight months with domain experts and stakeholders in systems biology, bioinformatics, and causal reasoning applied to the study of infectious diseases. Based on these interactions, we identified the following high-level design goals (DG):
\begin{itemize}
	\setlength\itemsep{0pt}
    \item \textbf{DG1. Provide scalable, interactive, and performant visualizations of biological graphs.} Such graphs range from small ones visualized in standard ways to those with over tens of thousands of nodes and hundreds of thousands of edges. The visualization approach should therefore be scalable with real-time interactivity, and should organize knowledge to be as visually digestible as possible.   
    Leveraging the data and ontology provided by our knowledge assembly collaborators \cite{EMMAA, INDRA}, our approach should enable biomedical researchers to query and analyze the high-level structures of biological graphs derived from domain literature, thereby gaining insight on the ontological structure of the extracted causal knowledge.
    \item \textbf{DG2. Provide iterative local analysis coordinated with global context.} While interactive graph overviews are useful to identify high-level structural and ontological patterns, biomedical researchers need to isolate and extract relevant subgraphs for more focused analysis. They also need to iterate on these subgraphs as needed while keeping global context in mind. For example, to judge drug side effects, our approach must enable researchers to identify incoming and outgoing pathways from particular agents, and iteratively expand or truncate the subgraph under analysis in light of all pathways available in the global graph. 
    \item \textbf{DG3. Promote scientific knowledge synthesis and discovery.} Researchers should be able to access backing scientific corpora to further contextualize their analysis and explore related knowledge spaces. Using data provided by our collaborators \cite{EMMAA, INDRA, cosmos}, our visualization approach allow researchers to seamlessly navigate back and forth from biological graphs to the scientific corpus at large. In this manner, researchers can trace pathways expressed in graph form to source documents and explore related knowledge to broaden their analysis. 
\end{itemize} 
\section{Data Processing}
Our approach leverages graph and knowledge data from collaborating systems in INDRA and COSMOS. INDRA (the Integrated Network and Dynamical Reasoning Assembler) assembles a network of biological processes from statements extracted from source documents using natural language processing techniques \cite{INDRA}. Extracted mechanistic and causal assertions are standardized and mapped to a biological ontology \cite{INDRA_BioOntology}, generating an assembled set of causal statements that constitute a biological graph . We further normalize this statement data into a multidigraph optimized for real-time browser rendering. Here, biological agents are mapped to nodes clustered as per their ontological category and causal statements are mapped to directed edges between agents, with edges bundled into hyper-edges to enable layout generation at each level of the ontology. 

COSMOS is a knowledge discovery platform that automates the process of extracting and assimilating heterogeneous artifacts from diverse scientific publications \cite{cosmos}. In addition to enabling search over a large corpus of publications, COSMOS extracts tables, text, images and source code. The corpus provided by COSMOS consists of 176,000 documents, complete with metadata and document artifact extraction. Knowledge search over this corpus is made available in our prototype via API. The corpus is also available as an embedding dataset, with each document cast as a vector embedding \cite{rehurek_lrec}. To these embeddings, we apply dimensional reduction via UMAP and hierarchical clustering via HDBSCAN to first project the embeddings into a 2D space and then group semantically proximate documents as nested clusters \cite{McInnes2017, mcinnes_umap_2020}. Moreover, for each cluster, we compute the alpha shape of its point subspace to establish a polygon cluster boundary \cite{edelsbrunner_shape_1983}.


\section{Visualization Design}
The following two subsections describe the two main spaces of our approach - the Graphs space and the Knowledge space - that were developed in light of the above design goals (DG). While the skeleton and core components of our approach are in place, we continue to refine the visualization, interaction, and analytic design for both spaces with domain experts.  

\subsection{Graphs Space}
The Graphs space is composed of two coordinated views (Fig. \ref{fig:teaser}) that seek to simultaneously provide the global and local perspectives needed for multi-scale graph sense-making \cite{pienta_scalable_2015}. As the default view, the \textit{Global View} (Fig. \ref{fig:teaser}. A) provides a visual overview of a biological graph, and has multi-faceted search, real-time navigation, and semantic zoom. The \textit{Local View} (Fig. \ref{fig:teaser}. B) is a sandbox to which search results can be iteratively added as a flow graph for further inspection. Our coordinated approach uses linked filtering and highlighting so changes in one view are reflected in the other. In this way, we look to enable biomedical researchers to engage in local subgraph analysis while contextualizing this analysis within the global context of the large-scale graph. 

\subsubsection{Global View}
The Global View shows an overview of a selected biological graph [\textbf{DG1}] assembled from large sets of biological causal statements. Each causal statement represents regulations between biological agents such as proteins or viruses (e.g. CDC12 phosporylates MID1). The overview global graph displays nested clusters of biological agents organized in the background biomedical ontology. 
We use a hierarchical circle-packing graph layout\cite{d3-hierarchy_2021} to show high-level relationship structure and the nested concepts of the ontology [\textbf{DG1}]. Each ontological group is rendered as rings (in blue) around their children (in orange). To avoid displaying hundreds of thousands of edges, we also apply edge bundling techniques\cite{holten_hierarchical_2006}, where edges with multiple nodes grounded to the same ontological category are bundled into hyper-edges. Hyper-edges are also wrapped around ontological rings to avoid edge clutter between groups, and hyper-edge brightness corresponds to the number of bundled statements. Semantic zooming progressively discloses the nested categories of the ontology as the researcher dives deeper into an ontological branch. 

This view also allows researchers to interrogate biological graphs by using multi-faceted queries [\textbf{DG2}]. Researchers can chain multiple queries related to node attributes (e.g. node degree), edge attributes (e.g. edge type), and path queries via a search box at the top of the Graphs space (Fig. \ref{fig:teaser}). Query results are displayed on the Global View using contrast-based highlighting, with results foregrounded and the rest of the graph faded out.

\subsubsection{Local View}
While such global views provide high-level structural information, they face sense-making challenges for multi-scale graphs\cite{keim_visual_2001, holten_hierarchical_2006, pienta_scalable_2015}. To this end, we coordinate the Global View with the Local View (Fig. \ref{fig:teaser}.B), where the latter displays a subgraph containing the subset of causal statements resulting from preceding search queries [\textbf{DG2}]. 

Edges are visually encoded based on the implied polarity of the regulation type (e.g. activation, inhibition), curation state (human-verified or unexamined), and directionality (directed or not). We use the following scale to encode edge polarity: blue for ``positive'', red for ``negative'', and gray for ``unknown.'' We use a filled arrow for directed edges and a filled box arrow for undirected edges. Human-verified statements are shown with a filled circle, and unexamined statements have empty circles (incorrect statements are removed during data pre-processing) (Fig. \ref{fig:teaser}. B). As directionality is critical for understanding causality \cite{Wright:2018}, we use a traditional left-to-right flow layout algorithm\cite{Sugiyama:1981} with recent extensions \cite{elkjs:2020}. 

Selecting a node in the Local View opens a Drill-down Panel for the Graphs Space to show node metadata (e.g. agent description) and link suggestions for incoming and outgoing relationships ranked by supporting evidence [\textbf{DG2}]. Neighborhood suggestions are also highlighted in the Global View so researchers can visualize the global connectivity of a given node. Selecting one or more of these relationship suggestions adds them to the subgraph, which helps with neighborhood expansion. 

\subsubsection{Graphs to Knowledge}
To aid trust-building, researchers need to link their graph exploration back to scientific knowledge [\textbf{DG3}]. We build on the links present in the provided data. For example, selecting an edge reveals the underlying evidence extracted by INDRA, where this evidence is surfaced as a short-text fragment (Fig. \ref{fig:teaser}. C). Clicking the text opens a modal dialog displaying associated metadata on both source and neighborhood documents provided by COSMOS. The neighborhood of a given document is taken as the set of semantically similar documents within the corpus\cite{rehurek_lrec}. In this manner, researchers can seamlessly trace a graph edge to its underlying textual evidence and then to the backing scientific paper.

\subsection{Knowledge Space} 


The Knowledge space displays an interactive overview of the scientific corpus. The corpus data contains all source material from which all biological graphs have been assembled, thus encouraging moving back and forth between graph-centric analysis and the literature at large [\textbf{DG3}]. Two perspectives on the knowledge corpus [\textbf{DG3}] are available via: a \textit{Card-based View} and a \textit{Clusters View}.


\subsubsection{Card-based View}
The Card-based View has a tabular layout with each document as a card, including a preview image of a document artifact and title. Cards organize information in chunks, which aids in user scannability. As in the Graphs space, the Knowledge space also has rich search capabilities where researchers can perform multiple compound queries on document attributes, including free text, author, publisher, and the presence of artifacts such as tables or figures. 

\subsubsection{Clusters View}

\begin{wrapfigure}{r}{\textwidth}
    \includegraphics[width=\linewidth] {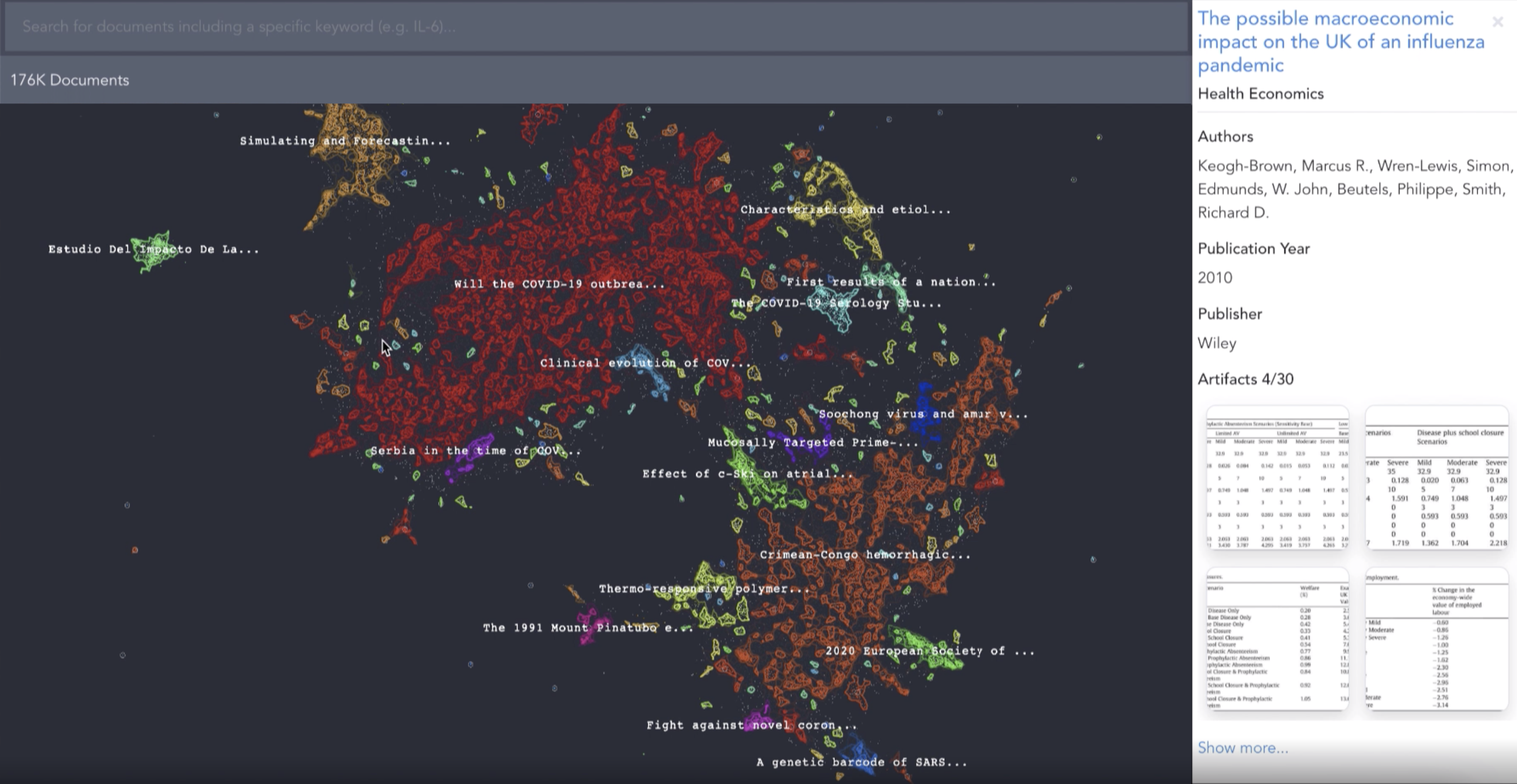}
    \centering
    \caption{The Clusters View represents the document corpus as an interactive 2D topology, where cluster members (documents) have the same color and are enclosed in a bounding polygon.}
    \label{fig:knowledge}
\end{wrapfigure}

While the Card-based View supports visual scannability, it suffers from visual scalability for large numbers of documents. The Clusters View (Fig. \ref{fig:knowledge}) thus provides a more scalable perspective on the scientific corpus using document embeddings. Organized in a 2D topology, a point in this visual space is a document, and spatial distance between points measures semantic similarity. Cluster members have the same color and are enclosed within a bounding alpha shape. Documents designated as ``noise'' by the clustering algorithm (not in any cluster) are coloured grey with no boundary. To visually encode cluster hierarchy, color hue is preserved across levels with lower opacity being applied to coarsest clustering levels. 




\subsubsection{Metadata and Knowledge to Graphs}
Selecting a document from either a card in the Card-based View or a point in the Clusters View opens a Drill-down Panel in the Knowledge Space with three tabs for deeper interrogation of the document content. The \textit{Preview} tab shows document metadata including title, DOI, authors and publisher-related information, with further drill-down available via a dialog with more detailed information including figures and text excerpts as well as the link back to the source document. This allows the researcher to discover related documents in addition to using spatially proximate documents in the Cluster view. The \textit{Graphs} tab contains links to existing biological graphs in the Graphs space, and fortifies the connection between user analysis and background scientific literature. The \textit{Entities} tab shows related keywords that helps researchers guide subsequent search. 
\section{Usage Scenario} \label{sec:scenario}

While our approach remains general across various biomedical graph-based use cases, we focus here on a usage scenario relevant to our current era around the exploration of biological mechanisms of COVID-19. A biomedical researcher is investigating potential drug treatments for SARS-CoV-2. Due to prior knowledge, they know that SARS-CoV-2 leads to the release of cytokines like IL6, which is associated with more severe symptoms and higher mortality in COVID-19 patients \cite{costela-ruiz_sars-cov-2_2020}. They hypothesize that identifying inhibitors of IL6 may provide potential therapeutics for the treatment of SARS-CoV-2. 

They start their investigation by selecting the COVID-19 graph from the available list in the Graphs space. Using their prior knowledge of articles related to cytokine release syndrome in severe COVID-19 cases \cite{zhang_cytokine_2020} and recommended medications for COVID-19 \cite{samaee_tocilizumab_2020, ma_study_2020}, they do a search query using the DOIs of these papers. Highlighted results in the Global View show the causal relationships extracted from these papers (Fig. \ref{fig:teaser}. A). To further examine these results, they click on the ``Open Local View'' button, which displays the corresponding subgraph in the Local View (Fig. \ref{fig:teaser}. B). They then select the IL6 node to open the Drill-down Panel with node metadata  and neighboring relationships. 
They explore the incoming relationships for IL6 and identifies a relationship with SARS-CoV-2, and adds this relationship to the subgraph. They then click on the new relationship to read the underlying evidence, which confirms their prior knowledge i.e. that SARS-CoV-2 increases the amount of IL6.

They continue exploring the subgraph in the Local View and notices that the relationship from tocilizumab to IL6 is encoded in red, which indicates that tocilizumab is an inhibitor of IL-6. They start to wonder if treating a COVID-19 patient with tocilizumab alter their chances of survival.
To better understand the involved biological mechanisms, they click on the edge to read the pieces of evidence underlying this relationship in the Drill-down Panel (Fig. \ref{fig:teaser}. C). In this panel, they see a check-mark beside the relationship heading, indicating that this relationship has been vetted by domain experts and is also supported by 39 pieces of evidence. In this light, they execute a path query that is chained with their prior queries. The results show a pathway from tocilizumab to IL6 to COVID-19, suggesting that tocilizumab (insofar as it acts upon IL6 and therefore could inhibit the cytokine storm) might be a drug candidate for severe COVID-19. Having formulated a hypothesis around tocilizumab and COVID-19 survival rates, they now want to identify if tocilizumab could have any side effects. To explore this, they inspect the 121 outgoing nodes from tocilizumab and see a relationship to immune responses. After adding this relationship to the subgraph, they find that it is of the inhibition type. By inspecting the underlying evidence, they also learn that tocilizumab may lower the ability of the immune system, increasing the risk of superinfections \cite{somers_tocilizumab_2020}. 
Through this quick exploration using our approach, the researcher has been able to quickly identify a potential drug to treat COVID-19 as well as potential side effects. They can continue to expand this line of inquiry to the backing scientific corpus available in the Knowledge space, and look to discover publications and research artifacts related to this drug. They can also choose to explore other biomedical hypotheses across the COVID19 graph, and quickly isolate relevant graph structures and textual support to further analysis.
\section{Expert Evaluation}

For an initial assessment on our approach to inform future iteration, we conducted a two-part evaluation: (i) a focus group with five external researchers for high-level feedback; and (ii) an online questionnaire focused on usability and utility with one biomedical researcher with over ten years of experience. All experts conduct research on systems biology, bioinformatics, and causal reasoning using biological models in their daily work.
For the focus group, we presented a demo of the tool before the experts were asked to give general feedback about system function in a group discussion that lasted around two hours. 
For the questionnaire, the expert was first asked to complete a task scenario, which was the same as the usage scenario described in Section \ref{sec:scenario} for investigating potential drug treatments for SARS-CoV-2.
After completing the scenario, the expert was encouraged to keep using the tool, then respond to the questionnaire.

The overall opinion of the domain experts was quite positive regarding the ease of interaction, the scale covered and the overall usefulness of the prototype. 
Specifically, the experts found the coordination between the Global and Local Views very useful for navigating the contained knowledge. 
The experts felt that this prototype achieves the goal of supporting query formulation while representing biological relationships in a visually organized way. Some critical feedback received regarded a slight delay when conducting large-scale path queries (due to the query's time-complexity not currently optimized over large graphs), inconsistencies in some path query results (because of path alternatives), and the general need for better notifications. We plan to address all these concerns in the near future. Overall, domain experts agreed that they would like to use the tool frequently as new features, optimizations and improvements continue to be developed.

\section{Conclusion and Future Work}

In this paper, we introduced a visual analytics approach for scalable exploration of biomedical knowledge that can aid across a range of biomedical use cases, from disease propagation to drug discovery. To demonstrate the feasibility of our approach, we implemented a web-based prototype that  displays 16 biological models against a corpus of 176,000 documents. Positive feedback from domain experts underscores the usefulness and usability of our prototype in helping them to explore, formulate and validate hypotheses. 

Several promising directions exist in terms of future research. While the Global View is visually scalable for graphs of thousands of nodes and edges, better overviews can be provided to the user via high-level summaries and aggregate graph statistics that surface interactively during graph exploration. Search remains a critical research thread, given the scale and complexity of biological graphs. Highlighting and navigation of search results could be further improved. Search functionality can also better incorporate user context, so more relevant results are returned.
Similarly, refinements in the Local View can improve neighborhood suggestions for user-extracted subgraphs. For example, there are almost 2,000 incoming nodes and 1,000 outgoing ones for IL-6 alone in the COVID-19 graph: path ranking based on user context could help prioritize the edge list for the problem at hand. For such global and local improvements, we look to explore graph embedding techniques that convert graph structures into vector representations. Such representations enable various downstream analytic tasks, including similarity search for pathways and subgraphs, path ranking, link prediction and so on. In the Knowledge space, we plan to refine our hierarchical clustering for better visual separation, as well as enrich the view with search capability. We also plan to incorporate citation graphs and add semantic layers such as topics and extracted artifacts to better facilitate exploration across the available corpus.

On the whole, the generality of our approach means it can extend to other workflows that utilize hierarchically organized knowledge graphs that are linked to backing documents. In this way, we believe our approach can be applied to a range of biomedical use-cases and ultimately support knowledge discovery across different scientific domains.

\acknowledgments{
This work was supported by the Defense Advanced Research Projects Agency (DARPA) under Contract (HR00111990005). 
}

\bibliographystyle{abbrv-doi}

\bibliography{template}

\newpage
\appendix
\section{Appendix}


\end{document}